\documentclass[aps,preprint,amsmath,amssymb]{revtex4}
\usepackage{graphicx,amssymb,axodraw}

\begin{document}

\title
{\Large \bf Same-sign  single dilepton productions at the LHC}

\author{
Chian-Shu~Chen\footnote{e-mail: chianshu@gmail.com},
Chao-Qiang~Geng\footnote{e-mail: geng@phys.nthu.edu.tw}, and
Dmitry~V.~Zhuridov\footnote{e-mail: zhuridov@phys.nthu.edu.tw}}
\affiliation{Department of Physics, National Tsing Hua University,
Hsinchu, Taiwan 300}

\date{\today}


\begin{abstract}
We examine  the same-sign  single dilepton productions of
$\ell_i^{\pm}\ell_j^{\pm}\ (\ell_{i,j}=e,\mu)$ in high-energy
proton-proton collisions at the Large Hadron Collider (LHC) in
models with doubly charged Higgs scalars as well as heavy Majorana
neutrinos. We demonstrate  that these spectacular productions can
be detected at the LHC for a class model in which the doubly
charged Higgs scalars couple only to the right-handed charged
leptons. The ranges of the possible doubly charged Higgs masses
and mixings to observe the processes at the LHC are discussed.
\end{abstract}

\maketitle

It is well accepted  that neutrinos have very small masses.  However, the origin of such smallness remains unclear.
One of the most popular solutions is that they arise from the seesaw mechanism with one or
 more  right-handed heavy Majorama neutrinos (HMN). On the other hand,
without right-handed neutrinos, it is well known that  the simplest way
to have a Majorana neutrino mass term  in the standard model (SM)
is to introduce a  complex
triplet Higgs $T$ with the hypercharge of $Y=-2$, defined by
\begin{eqnarray}
T=\left(\begin{array}{cc}T^0 & \frac{T^-}{\sqrt{2}} \\\frac{T^-}{\sqrt{2}} & T^{--}\end{array}\right),
\end{eqnarray}
which can couple to  $SU(2)_L$ lepton doublets $(L_{iL})$
\cite{Higgs}
\begin{eqnarray}
\label{LLT}
    {\cal L}_L=g_{ij}\overline{{L}_{iL}^c}T^{\dag}L_{jL}+{\rm H.c.},
\end{eqnarray}
where $g_{ij}$ are the coupling constants, $i,j=e,\mu,\tau$ and $c$ stands for the
charge conjugation.
The neutrino masses are generated to be $g_{ij}v_T$ after the triplet scalar $T$ receives
 the vacuum expectation value (VEV) of $v_T$.
Since the major goals of  the Large Hadron Collider (LHC) are
searching for Higgs scalars and understanding the mechanism of the
fermion mass generation, the HMNs and the triplet Higgs should be
parts of the studies at the LHC.

The most  interesting  models which contain the triplet are
left-right symmetric and little Higgs models \cite{LR,Little,Han}.
Phenomenologically, the doubly charged scalar in the complex
triplet could decay into the like-sign dileptons
($T^{\pm\pm}\rightarrow \ell^{\pm}_i\ell^{\pm}_j$) with a high
invariant mass, which provides a spectacular  signature with a
relatively small background \cite{QCDbackground} at hardron
colliders. A current limit set by the direct search at the
Tevatron in Fermilab  is $M_{T^{\pm\pm}}>136$~GeV \cite{Tevatron},
in which the Drell-Yan (DY) annihilation processes
$q\bar{q}\rightarrow \gamma^*,Z^*\rightarrow T^{++}T^{--}$ to the
final states of $e^{\pm}e^{\pm},e^
{\pm}\mu^{\pm},\mu^{\pm}\mu^{\pm}$ were used and a long-lived
doubly-charged scalar, corresponding to $g_{ij}\gtrsim10^{-5}$,
was assumed. However, from the current  neutrino mass upper bounds
\cite{PDG} of $0.1$ eV, extracted from the neutrino oscillation
data and cosmological experiments, the coupling of $g_{ij}$ cannot
be large if $v_T$ is not too small. Note that $v_T\le 4.41$~GeV
\cite{CGN,Chen} is constrained by the precision data of
$\rho=1.002^{+.0007}_{-.0009}$ \cite{PDG}.

Accordingly, one concludes that in the model with ${\cal L}_L$  of
Eq. (\ref{LLT}), the  production of $T^{\pm\pm}$ in the $W$-boson
fusion decaying into a like-sign single dilepton,
$W^{\pm}W^{\pm}\rightarrow T^{\pm\pm}\rightarrow
\ell^{\pm}_i\ell^{\pm}_j$, are too small to be found generally at
the LHC due to the following reasons: (a) the production rates are
proportional to $(v_T/v)^2$ which is numerically small even  $v_T$
is set to be close to the upper limit; and (b) as
$g_{ij}\lesssim10^{-10}$ for $v_T\sim 4.41$~GeV, the widths of
$T^{\pm\pm}\rightarrow \ell^{\pm}_i\ell^{\pm}_j$ are very small
and other channels would be opened to dominate over these
dileptons signatures \cite{HW}. Moreover, small coupling constants
$g_{ij}$ are needed in order to fit the neutrino mixing matrix.

Recently, a model was proposed \cite{CGN} with the $SU(2)_L$
complex triplet $T_{(-2)}$ and an additional doubly-charged
singlet $\Psi_{(4)}$  to the SM, where the subscript denotes the
hypercharge. In the model, a new Yukawa interaction, involving
$SU(2)_L$ charged lepton singlets $(\ell_R)$,
\begin{eqnarray}
\label{RRT}
 {\cal L}_R&=& Y_{ij}\overline{{\ell}^c_{iR}}{\ell}_{jR}\Psi +{\rm H.c.},
 \end{eqnarray}
is introduced due to $\Psi$ but the one in Eq. (\ref{LLT}) is
forbidden by imposing some symmetry for the Higgs fields such as
\begin{eqnarray}
\phi\rightarrow+\phi,\ \phi'\rightarrow-\phi',\ T\rightarrow-T \
{\rm and}\ \Psi\rightarrow+\Psi, \label{DS}
\end{eqnarray}
where  an extra Higgs doublet $\phi'$ has been also included.
However, since  the extra doublet leads to no new effects
\cite{CGZ} on the fermion couplings, the structure of the
doubly-charged Higgs scalars as well as the phenomenology in Refs.
\cite{CGN,Chen}, we will not discuss it further here.
In this model, as the neutrino masses are generated radiatively at
two-loop level \cite{CGN,Chen,Zee}, the small neutrino mass problem can be
naturally understood even with $Y_{ij}=O(1)$ and $v_T$ around the
upper limit simultaneously.

In this paper, we concentrate on doubly-charged scalars of
$T^{\pm\pm}$ and $\Psi^{\pm\pm}$. The two fields can form
doubly-charged massive physical states $P_1^{\pm\pm}$ and
$P_2^{\pm\pm}$ with the mixing angle $\delta$. It was argued in
Ref. \cite{Chen} that at least one of the doubly charged Higgs
scalars is well within the reach of the LHC. We take $P_1$ to be
this (lighter) state and focus on its phenomenology. In
particular, we investigate the processes $pp\rightarrow
P_1^{\pm\pm}X\rightarrow\ell_i^\pm \ell_j^{\pm}X$ under the
conditions of the LHC: $\sqrt{s}$=14 TeV and $L$=320~fb$^{-1}$,
where $\sqrt{s}$ is the beam energy and $L$ is the integrated
luminosity per year. We will also consider the contributions to
the processes  due to the HMNs. We will choose the condition
\begin{eqnarray}\label{sL}
    \sigma L\geq n
\end{eqnarray}
as $n$ events of the observation criteria for the process, where
$\sigma$ denotes the cross section.

We start by evaluating the differential cross sections for the
processes
\begin{equation}
\label{pp}
 pp\rightarrow\ell_i^\pm\ell_j^{\pm}X
\end{equation}
at the LHC  via the intermediate doubly charged Higgs
scalar $P^{\pm\pm}_{1}$
by neglecting the transverse polarizations of
$W$ bosons and quark mixings, where $X$ represents
2 jets, denoted as $JJ$.
The leading-order Feynman diagram for
the processes in Eq. (\ref{pp}) is shown in Fig.~1.
For $\ell_{i,j}=e$ or $\mu$, one has spectacular signatures of the same-sign dilepton pairs of
$e^{\pm}e^{\pm}$ or $\mu^{\pm}\mu^{\pm}$ or $e^{\pm}\mu^{\pm}$ without missing energy.
For the modes with one or two $\tau$ leptons, the final states are the above dilepton pairs but with missing energy
or pions with missing energy.
In this study, we shall not discuss  the productions with missing energy as they are suppressed.
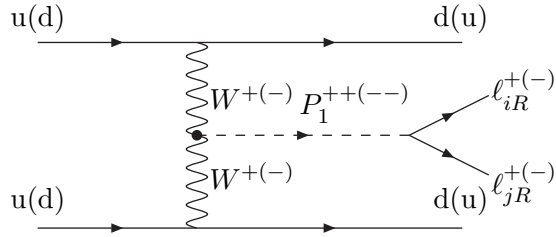
\begin{figure}[h]
\begin{picture}(200,100)(0,0)
\ArrowLine(10,80)(70,80) \ArrowLine(70,80)(170,80)
\Text(10,95)[t]{u(d)} \Text(170,95)[t]{d(u)}
\Photon(70,80)(70,45){4}{5} \Text(75,60)[l]{$W^{+(-)}$}
\Vertex(70,45){2} \Photon(70,10)(70,45){4}{5}
\Text(75,30)[l]{$W^{+(-)}$} \DashArrowLine(70,45)(150,45){4}
\Text(130,50)[b]{$P_1^{++(--)}$} \ArrowLine(10,10)(70,10)
\ArrowLine(70,10)(170,10) \Text(10,15)[b]{u(d)}
\Text(170,15)[b]{d(u)} \ArrowLine(150,45)(180,60)
\Text(194,55)[b]{$\ell^{+(-)}_{iR}$} \ArrowLine(150,45)(180,30)
\Text(194,35)[t]{$\ell^{+(-)}_{jR}$}
\end{picture}
\caption{Feynman diagram for
$pp\rightarrow\ell_i^\pm\ell_j^{\pm}JJ$ mediated by
$P_1^{\pm\pm}$.}\end{figure}
According to Ref. \cite{Chen}, the gauge-scalar and the lepton-scalar couplings are given by
\begin{eqnarray}
\label{gauge-scalar} \frac{g^2}{\sqrt{2}}v_Tc_\delta
    W_\mu^+W_\nu^+P_1^{--}+{\rm H.c.}\ {\rm and}\
    Y_{ij}s_\delta P_1^{--}\overline{\ell_{iR}^c} \ell_{jR}
    + {\rm H.c.},
\end{eqnarray}
respectively, where $c_\delta\equiv\cos\delta$ and $s_\delta\equiv\sin\delta$.
The decay of $P_1^{\pm\pm}$ can proceed by four types of channels:
$P_1^{\pm\pm}\rightarrow \ell_{iR}^{\pm}\ell_{jR}^{\pm}$,
$P_1^{\pm\pm} \rightarrow W^{\pm}W^{\pm}$,
$P_1^{\pm\pm} \rightarrow W^{\pm}P^{\pm}$
and
$P_1^{\pm\pm}\rightarrow W^{\pm}W^{\pm}T^0_a$, where $P^{\pm}$
 and $T_a^0$ are the  single-charged and neutral components of the Higgs scalars
in the model, respectively. The decay widths are given by \cite{Chen}
\begin{eqnarray}
\Gamma(\ell_{iR}^{\pm}\ell_{jR}^{\pm})&=&(1+\delta_{ij})\frac{|Y_{ij}|^2}{16\pi}s^2_{\delta}M_{P_1}, \\
\Gamma(W^{\pm}W^{\pm})&=&\frac{g^4v_T^2c^2_{\delta}}{16\pi M_{P_1}}
\sqrt{1-\frac{4M_W^2}{M^2_{P_1}}}\left(3-\frac{M_{P_1}^2}{M_W^2}+\frac{M^2_{P_1}}{4M_W^2}\right), \\
\Gamma(W^{\pm}P^{\pm})&=&\frac{g^2c^2_{\delta}M^3_{P_1}}{16\pi
M_W^2}\lambda^{3/2}
\left(1,\frac{M_W^2}{M_{P_1}^2},\frac{M_P^2}{M_{P_1}^2}\right),
\end{eqnarray}
where $\lambda(x,y,z)=x^2+y^2+z^2-2xy-2xz-2yz$. The three-body
decay modes are expected to be relatively suppressed by the phase
space compared to the two-body ones. In  Fig.~2, we show the decay widths
with two extreme
 cases of the mixing angles.
 The like-sign dilepton decays provide  clean and almost negligible
SM background signatures. Moreover, the branching ratios depend on
the Yukawa couplings $g_{ij}$ ($Y_{ij}$ in our case) which are
strongly linked with the different scenarios for  the neutrino mass generation mechanisms ~\cite{HW,Garayoa,NeutrinoMass}.
\begin{figure}[ht]
  \centering
    \includegraphics*[width=2.8in]{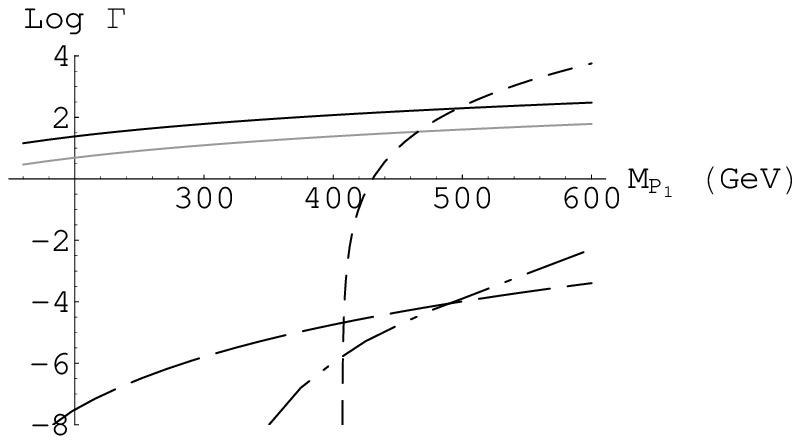}
     \includegraphics*[width=2.8in]{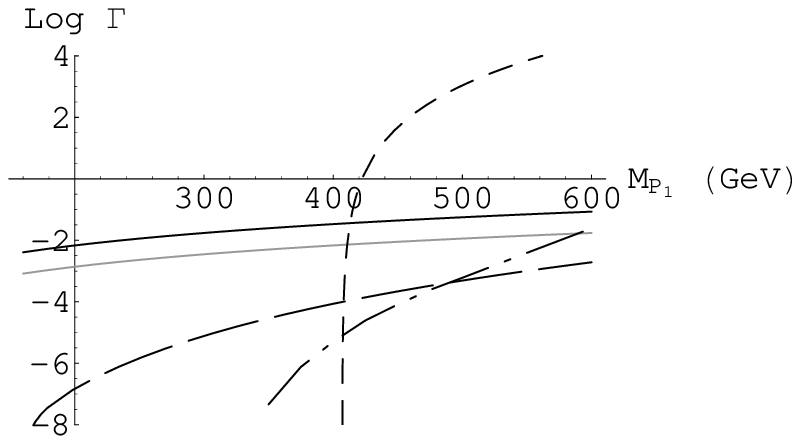}
  \caption{Logarithms of the decay widths (in units of GeV) of $P_1^{\pm\pm}$  as functions of $M_{P_1}$,
  where the left  (right) figure corresponds to
  the maximal (small) mixing of
  $\sin\delta=1/\sqrt{2}$ ($0.12$),
while the solid, dotted, long-dashed, short-dashed, and dot-dashed lines represent
   $\ell_i^\pm\ell_i^\pm$,
  $\ell_i^\pm\ell_j^\pm$ ($i\neq j$), $W^\pm W^\pm$, $W^\pm P^\pm$, $W^\pm W^\pm T_a^0$ modes,
  respectively.}
\end{figure}

The differential cross sections for the processes in Eq.
(\ref{pp}) are found to be
\begin{eqnarray}
\frac{d\sigma_\pm^{pp}}{d\cos\theta}
=A\left(\lambda_1^{ij}\right)^2 H_\pm^{pp}\,,
\label{cs}
\end{eqnarray}
where $\theta$ is the angle between the directions for $WW$ or
$qq$ and same sign leptons,
\begin{eqnarray}\label{c}
A&=&\frac{ G_{F}^{4}M_{W}^{6}}{2^7\pi ^{5}} =50~{\rm ab},\
    \lambda_1^{ij}= \sqrt{2-\delta_{ij}}
    \left|Y_{ij}\right| c_\delta s_\delta,
\nonumber\\
H_\pm^{pp}& =& \left(\frac{v_T}{M_W}\right)^2\int_{z_{0}}^{1}\frac{dz}{z}\int_{z}^{1}\frac{dy}{y}\int_{y}^{1}\frac{dx%
}{x}p_\pm\left( x,xs\right) p_\pm\left(
\frac{y}{x},\frac{y}{x}s\right) l\left( \frac{z}{y}\right) h\left(
\frac{s}{M_{P_1}^2}z\right), \label{F}
\end{eqnarray}
with
$z_{0}=M_{P_1}^{2}/s$.
In Eq. (\ref{c}), $h( t)$ are the normalized cross sections for the
subprocesses of $W^\pm W^\pm\rightarrow\ell_i^{\pm} \ell_j^{\pm}$, given by
\begin{eqnarray}\label{h}
h\left( t\right)= \frac{t\left( t-4M_W^2/M_{P_1}^2\right) }{\left(
t-1\right)^2 + \Gamma_{P_1}^2/M_{P_1}^2 }\,,
\end{eqnarray}
with the total decay width of $P_1^{\pm\pm}$ as:
\begin{eqnarray}
\Gamma_{P_1}=3\left[\Gamma(\ell_{iR}^{\pm}\ell_{iR}^{\pm})+\Gamma(\ell_{iR}^{\pm}\ell_{jR}^{\pm})_{i\neq
j}\right]+ \Gamma(W^{\pm}W^{\pm})+ \Gamma(W^{\pm}P^{\pm})+
\Gamma(W^{\pm}W^{\pm}T^0_a)\,,
\end{eqnarray}
$l(r)$ is the normalized luminosity (multiplied by $r$) of
$W^{\pm}W^{\pm}$ pairs in the two-quark system~\cite{Daw},
defined by
\begin{eqnarray}\label{l}
l\left( r\right) = -\left( 1+r\right) \ln r-2\left( 1-r\right)\,,
\end{eqnarray}
 and $p_{\pm}\left( x,Q^{2}\right)$ are the  quark  distributions in the proton,  which have the forms:
\begin{eqnarray}\label{p+}
p_+\left( x,Q^{2}\right) &=& x\sum_{i}q_{i}\left( x,Q^{2}\right)
=x\left(u+c+t+\bar{d}+\bar{s}+\bar{b}\right),\\
p_-\left( x,Q^{2}\right) &=& x\left(\bar u+\bar c+\bar
t+d+s+b\right) \label{p-}\,.
\end{eqnarray}
It is interesting to note that the angular distributions in Eq.
(\ref{cs}) are uniform on the quark level as there is only the $s$
channel diagram for each of the processes in Eq. (\ref{pp}).

In the numerical calculation of the differential cross sections in
Eq. (\ref{cs}), we use the CTEQ5 parton distributions \cite{cteq}
and take $\left|Y_{ij}\right|$~= 1, $s_\delta=
0.12$ or $1/\sqrt{2}$, and $v_T$=4~GeV \cite{Chen}.
We note that $|Y_{ij}|=1$ is
just a convenient choice.
The constraints on these couplings from the neutrino oscillation data,
rare decays, and $0\nu\beta\beta$ decays are studied in Ref.
\cite{Chen}, where it is shown that the weakest upper bound is for
the $\mu\mu$ production: $|Y_{\mu\mu}|<3.5$.
However, due to $Y_{\ell\tau}<0.2\ (\ell=e,\mu)$ and $|Y_{\tau\tau}|<0.02$ \cite{Chen} as well as
the small  branching ratios
of $\tau\to \ell\nu_{\tau}\bar{\nu}_{\ell}$,
 we shall exclude the  productions with one or two taus in our discussion.

In Fig.~3, we show the relation between the cross sections and the
mixing angle $\delta$ at $M_{P_1}=200$ GeV. Note that for $i\neq
j$ there is an additional factor 2 in Eq. (\ref{cs}). In the
figure, we also give the one event discovery limit (DL) at the LHC
according to the observation criteria in Eq. (\ref{sL}). We find
that since the decay widths depend on the mixing angle in
different ways, the maximal cross section is not happened in the
large mixing angle of $s_\delta=1/\sqrt{2}$ but around
$s_{\delta}\sim0.15$. In Fig.~4, we plot the cross sections of
$P_1^{\pm\pm}$ and the DL at various values of $M_{P_1}$. The rate
for $P_1^{++}$ is about twice to that of $P_1^{--}$ as expected
based on the larger u-quark content in the proton at the LHC. In
the case of $s_\delta=1/\sqrt{2}$, the processes via $P_1^{--}$
are unobservable at the LHC. On the other hand, in the case of
$s_\delta=0.12$, the cross sections drastically decrease as
$M_{P_1}$ is above 420 GeV because the decay channel of
$W^{\pm}P^{\pm}$ opens up and becomes a dominant mode as seen from
Fig.~2.
\begin{figure}[ht]
  \centering
    \includegraphics[width=0.4\textwidth]{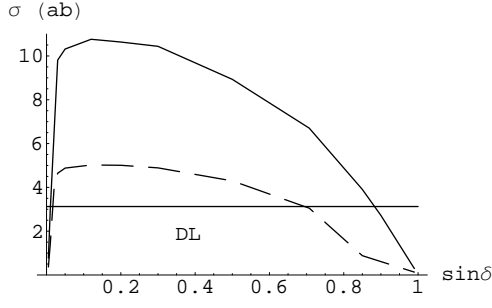}
  \caption{ Cross sections of $pp\rightarrow P_1^{++}X\rightarrow
  \ell_i^+\ell_j^+X$ (solid line) and $pp\rightarrow P_1^{--}X\rightarrow
  \ell_i^-\ell_j^-X$ (dashed line) as functions of  $\sin\delta$
at $M_{P_1}=200$~GeV, where the straight line is the one event
discovery limit (DL) at the LHC.}
\end{figure}
\begin{figure}[ht]
  \centering
    \includegraphics[width=0.5\textwidth]{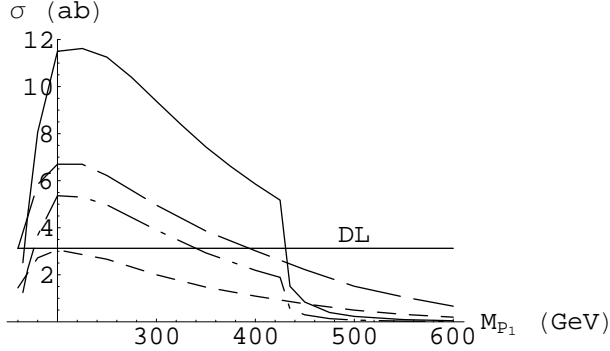}
  \caption{Cross sections of $pp\rightarrow P_1^{\pm\pm}X\rightarrow
  \ell_i^\pm\ell_j^\pm X$ as functions of $M_{P_1}$,
where the solid (dot-dashed) and long-dashed  (short-dashed) lines
stand for the processes $pp\rightarrow P_1^{++}X\rightarrow
  \ell_i^+\ell_j^+X$
($\ell_i^-\ell_j^-$) with $\sin\delta=0.12$ and $1/\sqrt{2}$,
respectively, while the straight line is the one event DL at the
LHC.}
\end{figure}
We can conclude that, at
the LHC, it is possible to detect  the processes in Eq. (\ref{pp}) via the intermediate
doubly-charged Higgs with its mass in the range from 180~GeV
to 400~GeV while the mixing is between $\sin\delta=0.03$ and
$0.85$. It should be pointed out that one may tune the
parameters to push the decay widths of $\Gamma(W^{\pm}P^{\pm})$ at higher $M_{P_1}$ to open
up the modes of  $W^{\pm}P^{\pm}$, then the searching range for
$P^{\pm\pm}_1$ may be extended.

We now study the 
 mechanism to produce dileptons
in $pp$ collisions  due to the intermediate
HMNs \cite{Almeida,Ali}. The processes in Eq. (\ref{pp}) mediated
by the HMN are illustrated in Fig.~5, where $M_{N_1}$ corresponds
to the mass of the lightest HMN. The differential cross sections
are given by
\begin{eqnarray}
\frac{d\sigma^{pp}_N}{d\cos\theta} =2A\left(\rho_1^{ij}\right)^2
N^{pp}\,, \label{csM}
\end{eqnarray}
with
\begin{eqnarray}
\rho_1^{ij}&=& \sqrt{2-\delta_{ij}} \left|u_{i1}u_{j1}\right|\,,
\nonumber\\
N^{pp} &=& \left(\frac{M_{N_1}}{M_W}\right)^2\int_{\tilde z_{0}}^{1}\frac{dz}
{z}\int_{z}^{1}\frac{dy}{y}\int_{y}^{1}\frac{dx%
}{x}p_\pm\left( x,xs\right) p_\pm\left(
\frac{y}{x},\frac{y}{x}s\right) l\left( \frac{z}{y}\right) n\left(
\frac{s}{M_{N_1}^2}z, \cos\theta\right), \label{FM}
\end{eqnarray}
where $\tilde z_{0}=4M_{N_1}^{2}/s$,  $u_{i1}$ are the
mixing matrix elements between the $i$th charged lepton and the  heavy neutrino, and
$n\left( t, \cos\theta\right)$ are
the normalized cross sections for the subprocesses $W^\pm W^\pm\rightarrow\ell_i^{\pm}
\ell ^{\pm}_j$, given by
\begin{eqnarray}\label{nM}
n\left( t, \cos\theta\right)=\left(
\frac{1-\cos\theta}{1-\cos\theta+2t^{-1}} +
\frac{1+\cos\theta}{1+\cos\theta+2t^{-1}} \right)^2.
\end{eqnarray}
Numerically, we find that these processes cannot be observed at
the LHC even with a lower $M_{N_1}$. However, it is still possible
at  some higher luminosity colliders beyond the LHC~\cite{Ali}. It
is interesting to note that
the angular distributions
in Eq. (\ref{csM}) given by the HMN mechanism
are not uniform in contrast with the
uniform ones in Eq. (\ref{cs}) by the doubly charged Higgs.
Moreover, the produced
same-sign leptons in Fig. 5 are  left-handed,
whereas  those are  right-handed
in Fig. 1.

\begin{center}
\begin{picture}(200,110)(0,0)
\ArrowLine(10,100)(80,100) \ArrowLine(80,100)(170,100)
\Text(10,95)[t]{u(d)} \Text(170,95)[t]{d(u)}
\Photon(80,100)(80,70){4}{5} \Text(85,85)[l]{$W^{+(-)}$}
\Line(80,70)(80,40) \Photon(80,10)(80,40){4}{5}
\Text(85,25)[l]{$W^{+(-)}$} \ArrowLine(10,10)(80,10)
\ArrowLine(80,10)(170,10) \Text(80,55)[]{$\times$}
\Text(85,55)[l]{$N_1$} \Text(10,15)[b]{u(d)}
\Text(170,15)[b]{d(u)} \ArrowLine(80,70)(150,70)
\ArrowLine(80,40)(150,40) \Text(155,70)[l]{$\ell^{+(-)}_{iL}$}
\Text(155,40)[l]{$\ell^{+(-)}_{jL}$}
\end{picture}
\\
{\small Fig.~5 Feynman diagram for the processes
$pp\rightarrow\ell^\pm_i\ell^{\pm}_jJJ$ mediated by a heavy
Majorana neutrino.}
\end{center}


In summary,
the possibility to observe the same-sign single dilepton  productions at the LHC has been
examined based on the doubly charged Higgs and Majorana neutrino mechanisms.
We have demonstrated
 that the productions of $pp\rightarrow\ell^\pm_i\ell^{\pm}_jJJ\ (\ell_{i,j}=e,\mu)$ can only be observed
at the LHC in the model with the doubly charged scalars coupling to the right-handed charged leptons.
In particular,  the ranges of  doubly charged Higgs masses and
mixings to observe the productions in terms of the one-event
discovery limits have been determined. We have  also shown  that
the angular distributions of the differential cross sections for
the processes are uniform on the quark level in contrast with the
non-uniform ones due to the Majorana neutrino exchange mechanism.
Finally, we remark that we have  considered the processes $e^\mp
p\rightarrow\ell^\mp_i\ell^{\mp}_jJJ$ via the intermediate
$P^{\mp\mp}_{1}$ and HMN. However, by using the same method as
for the $pp$ collisions we find that these processes cannot be
observed at the near future $ep$ colliders \cite{ABZ}.

\section*{Acknowledgements} This work is financially supported by
the National Science Council of Republic of China under the
contract \#: NSC-95-2112-M-007-059-MY3.


\end{document}